\documentstyle[prl,aps,floats]{revtex}
\input{psfig.tex}
\tighten
\begin{document} 
\draft
\twocolumn[\hsize\textwidth\columnwidth\hsize\csname@twocolumnfalse\endcsname 
\title{Phenomenology of a realistic accelerating universe using only Planck-scale physics}
\author{Andreas Albrecht, Constantinos Skordis}
\address{Department of Physics, UC Davis, One Shields Ave, Davis CA 95616}
\maketitle
\begin{abstract}

Modern data is showing increasing evidence that the Universe is
accelerating.  So far, all attempts to account for the acceleration
have required some fundamental dimensionless quantities to be extremely
small. We show how a class of scalar field
models (which may emerge naturally from
superstring theory) can account for acceleration which starts in the
present epoch with all the potential parameters $O(1)$ in Planck units.

\end{abstract}

\date{\today}

]

\renewcommand{\thefootnote}{\arabic{footnote}}
\setcounter{footnote}{0}

Current evidence that the Universe is accelerating\cite{BOPS},
if confirmed, requires dramatic changes in the field of theoretical
cosmology.   Until recently, there was strong prejudice against the
idea that the Universe could be accelerating.  There simply is no
compelling 
theoretical framework that could accommodate an accelerating universe.
Since the case for an accelerating universe continues to build (see
for example \cite{toco}), 
attempts have been made to improve the theoretical situation, with
some modest success.  Still, major open questions remain.

All attempts to account for acceleration
\cite{BBMSS,FHSW,CDF,CS,WS,CDS,HWDCS,SWZ,ZWS,MCBW,WCOS,TC,SS,MPR,PBM,ZS,KC,BBS,VL}
introduce a new type of matter (the ``dark energy'' or ``quintessence'')
with an equation of state $p_Q = w_Q \rho_Q$ relating pressure and
energy density.  Values of $w _Q < -0.6$ today are
preferred by the data\cite{PTW} and in many models $w_Q$ can vary over
time.  (In this framework, $  w_Q= -1, \dot w_Q =0$ gives a
cosmological constant.)  

One challenge faced by quintessence models is similar to the old ``flatness problem''
which is addressed by cosmic inflation.  Consider $\Omega_{tot} \equiv \rho_{tot}/\rho_{c}$ where
$\rho_{c}$ is the critical density (achieved by a perfectly flat
universe).  The dynamics of the standard big bang drive $\Omega_{tot}$
away from unity, and require extreme fine tuning of initial conditions
for $\Omega_{tot}$ to be as close to unity as it is today (inflation
can set up the required initial conditions). In models with quintessence
one must consider $\Omega_Q \equiv \rho_Q/\rho_c$ which is
observed to obey
\begin{equation}
\Omega_Q \approx \Omega_{other}
\equiv (\rho_{tot} - \rho_Q)/\rho_c
\label{OQcondition}
\end{equation}
 today.  The a ``fine tuning'' problem 
in quintessence models comes from the tendency for $\Omega_Q$ to
evolve away from $\Omega_{other}$. Equation
\ref{OQcondition} is achieved in these models either 1) by fine tuning
initial conditions or 2) by introducing a small scale into the
fundamental Lagrangian which causes $\rho_Q$ to only start the
acceleration today.  This
second category includes cosmological constant models and also a very
interesting category of ``tracker'' quintessence models\cite{SWZ,ZWS,ZS} which achieve
the right behavior {\em independently} of the initial conditions for the
quintessence field.  One then has to explain the small scale in the
Lagrangian, and this may indeed be possible \cite{PBBM}

Here we discuss a class of quintessence models which behave
differently.  Like the ``tracker'' models, these
models predict acceleration independently of the initial
conditions for the quintessence field.  These models also have
$\rho_Q(today)$ fixed by parameters in the fundamental
Lagrangian.  The difference is that all the parameters in our
quintessence potential are $O(1)$ in Planck units.  As with all
known quintessence models, our model does not solve the cosmological
constant problem:  We do not have an explanation for the location of
the zero-point of our potential.  This fact limits the
extent to which any quintessence model can claim to ``naturally''
explain an accelerating universe.  Recently
Steinhardt\cite{SL} has suggested that $M$-theory arguments specify
the zero-point of potentials in 3+1 dimensions.  Our zero point coincides
with the case favored by Steinhardt's argument.

We start by considering a homogeneous quintessence
field $\phi$ moving in a potential of the form
\begin{equation}
V(\phi) = e^{-\lambda \phi}.
\label{expV}
\end{equation}
We work in units where $M_P \equiv (8\pi
G)^{(-1/2)}= \hbar = c = 1$. The role of spatial variations in such a
field has been studied in \cite{CDF,CDS,MCBW,FJ,VL}. Inhomogeneities can be
neglected for our purpose, which is 
to study the large-scale evolution of the universe. We assume
inflation or some other mechanism has produced what is effectively a
flat Friedmann-Robertson-Walker universe and work entirely within that
framework. Cosmological fields with this type of exponential potential
have been studied for some time, and are well
understood\cite{H,B,CLW,CW,RP,BCC}. (A nice review can be found in \cite{FJ}.)

\begin{figure}[h]
\centerline{\psfig{file=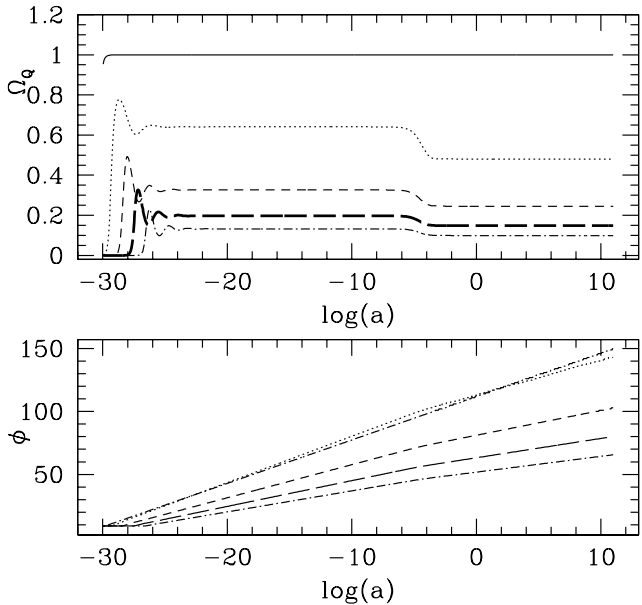,width=3.5in}}
\caption{The upper panel shows $\Omega_Q(a)$ for different constant values
 of $\lambda$. Each solution shows scaling behavior after the initial transient.  
The radiation-matter transition is evident at $a \approx 10^{-5}$. 
The heavy curve saturates a
generous interpretation of the nucleosynthesis bound, but still does
does not generate acceleration today.
The lower panel shows the evolution of $\phi(a)$ for the same
solutions. Note how $\phi$ varies very little while $a$ and $\rho_Q$
vary by many orders of
magnitude.  
 Today $a=1$, $a \approx 10^{-10}$ at nucleosynthesis, 
and $a\approx 10^{-30}$ at the Planck epoch.
The curves correspond to $\lambda = 1.5, 2.5, 3.5, 4.5,$ and $5.5$ (going
from top to bottom in the upper panel).}
\label{phiofa1}
\end{figure}

Let us review some key features: 
A quintessence field with this potential will approach a
``scaling'' solution, independent of initial conditions.  During 
scaling $\Omega_Q$ takes on a fixed value which depends only
on $\lambda$ (and changes during the radiation-matter transition).  In
general, if
the density of the dominant matter component scales as $\rho \propto
a^{-n}$ then after an initial transient the quintessence field obeys
$\Omega_\phi=\frac{n}{\lambda^2}$, effectively ``locking on'' to the
dominant matter component.  
Figure \ref{phiofa1}(upper panel)
shows $\Omega_Q(a)$ for scaling solutions in
exponential potential
models, where $a$ is the scale factor of the expanding universe
($a(today) =1$). At the Planck epoch $a \approx
10^{-30}$. In 
Fig. \ref{phiofa1}(upper panel) one can see the initial
transients which all approach the unique scaling solution determined
only by $\lambda$. In \cite{LS} it is shown that exponential potentials
are the {\em only } potentials that give this 
scaling behavior.

Scaling models are special because the condition 
$\Omega_Q \approx \Omega_{other}$ is achieved naturally 
through the scaling behavior.  The problem with these exponential
models is that no choice of $\lambda$ can give a model that
accelerates today {\em and} is consistent with other data.  The tightest
constraint comes from requiring that $\Omega_Q$ not be too large
during nucleosynthesis\cite{OSW} (at $a \approx 10^{-10}$).   The 
heavy curve in Fig \ref{phiofa1}(upper panel) just saturates a generous
$\Omega_Q < 0.2$ bound at nucleosynthesis, and
produces a sub-dominant $\Omega_Q$ today.  The combined effects
of sub-dominance and scaling cause $w_Q=0$ in the matter era, so this
solution is irrelevant to a Universe which is
accelerating today.

The lower panel of  Fig. \ref{phiofa1} shows how the value of $\phi$
changes by only about an order of magnitude over the entire history of
the universe, while the scale factor
(and $\rho_Q$) change by many orders of magnitude.  This effect, which is due
to the exponential form of the potential, plays a key role in
what follows.  The point is that modest
variations to the simple exponential form can produce much more
interesting solutions.  Because $\phi$ takes on values throughout
history that are $O(1)$ in Planck units, the parameters in the modified $V(\phi )$
can an also be $O(1)$ and produce solutions relevant to current observations.

\begin{figure}[h]
\centerline{\psfig{file=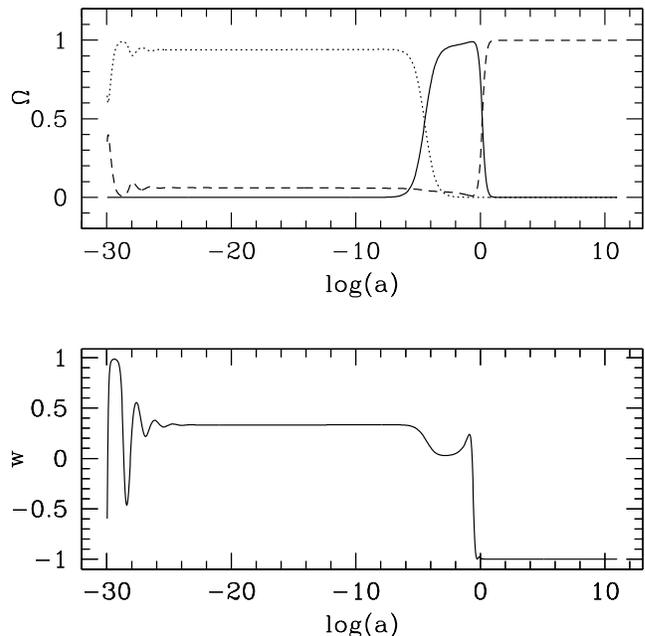,width=3.5in}}
\caption{Upper panel: A solution obtained by including a $V_p$ factor in
the potential. We show $\Omega_Q(a)$ (dashed), $\Omega_{\rm \it matter}(a)$
(solid) and $\Omega_{\rm \it radiation}(a)$ (dotted). The lower panel shows $w_Q(a)$. The solution shows the
normal radiation and matter dominated epochs and then an accelerating
quintessence dominated epoch at the end.  We have used $V_p$ given by
Eqn. \ref{Vp} 
with $B=34.8$, $\alpha = 2$, $A = .01$, and $\lambda = 8$.
Today $a=1$, $a \approx 
10^{-10}$ at nucleosynthesis, and $a\approx 10^{-30}$ at the Planck epoch.}
\label{goodone}
\end{figure}

Many theorists believe that fields with potentials of the form
\begin{equation}
V(\phi) = V_p(\phi ) e^{-\lambda \phi}.
\label{expVp}
\end{equation}
are predicted in the low energy limit of $M$-theory , where $V_p(\phi
)$ is a polynomial.  As a simple
example we consider  
\begin{equation}
V_p(\phi) = (\phi - B)^{\alpha} + A.
\label{Vp}
\end{equation}
For a variety of values for $\alpha$, $A$ and $B$ solutions 
like the one shown in Fig. \ref{goodone} can be produced. In this
solution $\Omega_Q$ is well below the nucleosynthesis bound, and the 
universe is accelerating today.  We show $\Omega_Q(a)$ (dashed),
$\Omega_{\rm \it matter}(a)$
(solid) and $\Omega_{\rm \it radiation}(a)$ (dotted). The lower panel in figure
\ref{goodone} plots $w_Q(a)$ and shows that the necessary negative
values are achieved at the present epoch.   

\begin{figure}[h]
\centerline{\psfig{file=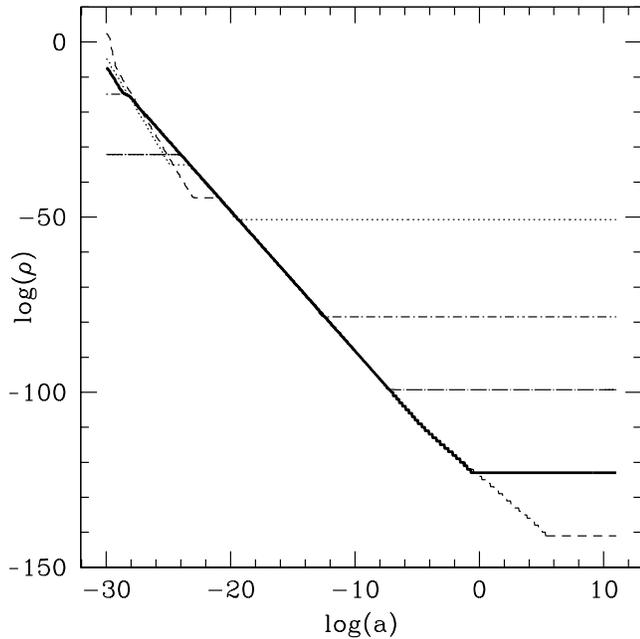,width=3.5in}}
\caption{Energy density ($\rho_Q$) {\em vs} $a$. The
heavy shows the solution from Fig
\ref{goodone}. This set of solutions shows that the point where
acceleration ($\rho_Q = const$) starts is controlled by Planck-scale
parameters in the Lagrangian. 
Moving from top to bottom
on the right side the values of $B$ (in Eqn. 4) are
14, 22, 28, 34.8, and 40.  We also varied the initial value of
$\phi$ between $0.3$ and $10$ to illustrate a range of initial 
transients.}  
\label{solutions}
\end{figure}

Figure \ref{solutions} illustrates how the solutions depend on the parameters
in $V_p(\phi )$.  We plot quintessence energy density $\rho_Q$ as a function of the
scale factor $a$.  After showing some initial transient behavior each
solution scales with the other matter for an extended period before
$\rho_Q$ comes 
to dominate.  The radiation-mater transition, which occurs at around
$a = 10^{-5}$ can be seen in figure \ref{solutions} as a change in the
slope in the scaling domain. The constant parameter $B$ in
Eqn. \ref{Vp}  has been selected from the range $14-40$ for these
models, yet the point of $\phi$ domination shifts clear across the
entire history of the universe.  In this picture, the fact that $\Omega_Q$ is
just approaching unity {\em today} rather than $10^{10}$ years ago is
put in by hand, as is the case with other models of cosmic
acceleration.  Our models are special because this can be
accomplished while keeping the parameters in the potential $O(1)$ in
Planck units.  Although we only illustrate the $B$ dependence
here, we have found that similar behavior holds when other parameters in
$V_p(\phi )$ are varied.

\begin{figure}[h]
\centerline{\psfig{file=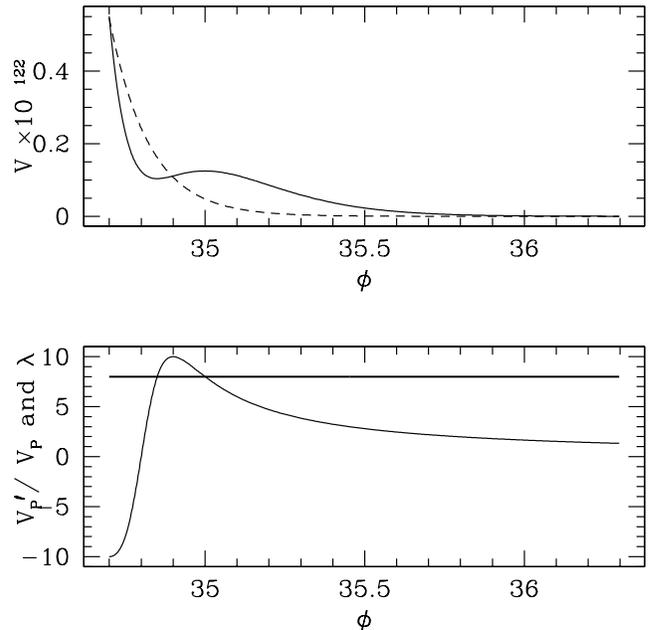,width=3.5in}}
\caption{The top panel shows a close-up of the interesting region
of $V(\phi)$ for the solution shown in Fig. \ref{goodone}. The dashed curve
show a pure $V(\phi) = \exp{(-\lambda \phi)}$ potential with $\lambda
= 8.113$ for comparison.  The the onset of acceleration is caused by
$\phi$ settling into the local minimum.  The fact that the feature
is introduced on such a small scale is due to the exponential factor
in the potential.  The lower panel shows
$V_p^{\prime}/V_p$ and 
$\lambda$ (the constant curve).  Where the two curves cross $V^{\prime}
= 0$.}  
\label{hump}
\end{figure}

Let us examine more closely what is going on:  The derivative of
$V(\phi )$ is given by
\begin{equation}
{ d V \over d \phi } = \left({V_p^{\prime} \over V_p} - \lambda \right)V.
\label{Vprime}
\end{equation}
In regions where $V_p$ is dominated by a single power-law  $\phi^n$ behavior
$V_p^{\prime} / V_p  \approx n/\phi $ which, unless $n$ is large, rapidly becomes $<< \lambda $
for values of $\lambda$ large enough to evade the nucleosynthesis
bound, leading to little difference from a simple exponential potential.  However, there will be points were $V_p$ can show other
behavior which {\em can} impact $V^{\prime}$.  Using equation \ref{Vp}
gives
\begin{equation}
{V_p^{\prime} \over V_p} = { \alpha (\phi - B)^{\alpha -1} \over (\phi - B)^{\alpha}
 + A }
\end{equation} 
This varies rapidly near $\phi = B$ and for $\alpha = 2$ peaks at a
value $V_p =
1/\sqrt{A}$. The upper panel of Fig. \ref{hump} shows the behavior of $V$ near $\phi =
B = 34.8 $ for the solution shown in Fig \ref{goodone}. The dashed curve shows a pure exponential
for comparison.  The lower panel
shows the curves ${V_p^{\prime} / V_p}$ and $\lambda$ (the constant
curve). Where these two curves cross $V^{\prime} = 0$. Because
the peak value $1/\sqrt{A} > \lambda $, two zeros are produced in
$V^{\prime}$ creating the bump shown in the figure.  In our solution $\rho_Q$ is coming to dominate near
$\phi = B$ because the field is getting trapped in the local minimum.
The behavior of the scaling solution ensures that $\phi$ gets stuck in
the minimum rather than rolling on through (regardless of the initial
conditions).  

At one stage in this work we focused on potentials of the form $V(\phi
) = \exp(-\lambda_{\rm \it eff} \phi)$ with the idea that $\lambda_{\rm \it eff}$ might not be
absolutely constant, but could be slowly varying with $\phi$.  We
considered forms such as $\lambda_{\rm \it eff} = \lambda(1 -
(\phi/B)^{\alpha})$ and found many interesting solutions, especially
for moderately large values of $B$ which make $\lambda_{\rm \it eff}$ slowly
varying.  For example, $\lambda =13$, $B = 65$, and $\alpha=1.5$ give
a solution similar to Fig \ref{goodone}.  If this form for $\lambda_{\rm \it eff}$ were taken
seriously for large $\phi$, then these models have an absolute minimum
in $V$ which $\phi$ settles into (or at least approached) at the start
of acceleration.  But our expression may just represent an
approximation to $\lambda_{\rm \it eff}$ over the relevant (finite) range of
$\phi$ values.  Of course we always can re-write Eqn. \ref{expVp} in
terms of $\lambda_{\rm \it eff}$ with $\lambda_{\rm \it eff} = \lambda -ln(V_p)/\phi$.
In the end we focused on potentials in the form of Eqn. \ref{expVp} because
they seem more likely to connect with ideas from $M$-theory.  Whatever
form one considers for $V$, the concept remains the same. Simple
corrections to pure $V= \exp{(-\lambda \phi)}$ can produce interesting
solutions with all parameters $O(1)$ in Planck units.

We should acknowledge that we use $O(1)$ rather loosely here.  In the
face of the sort of numbers required by other quintessence models or
for, say, a straight cosmological constant ($\rho_{\Lambda} \approx
10^{-120}$) numbers like $.01$ and $34.8$ are $O(1)$.  Also,
the whole ``quintessence'' idea has several important
open questions.  Some authors argue\cite{KL} that values of $\phi > 1
$ should not be considered without a full quantum gravitational
treatment, although currently most cosmologists do not to worry as long as
the {\em densities} are $<< 1$ (a condition our models easily
meet). Another issue that has been emphasized by Carroll \cite{SC} is that even
with the (standard) assumption that $\phi$ is only coupled to other
matter via gravity, there still will be other observable consequences
that will constrain quintessence models and require small couplings.
Because in our models $\dot \phi \approx 0$ today the tightest
constraints in \cite{SC} are evaded, but there would still be
effective dimensionless
parameters $\approx 10^{-4}$ required.

Looking toward the bigger picture, a general polynomial
$V_p$ will produce other features of the sort we have
noted.  Some bumps in the potential can be ``rolled'' over classically,
but may produce features in the perturbation spectrum or
other observable effects.  We are investigating a variety of
cosmological scenarios with a more general version of
$V_p$.  We are also looking at the effect of quantum decay processes
which are relevant to local minima of the sort we consider here.  We
expect a range of possibilities depending on the nature of $V_p$.

In conclusion, we have exhibited a class of quintessence models which
show realistic accelerating solutions. These solutions are
produced with parameters in the quintessence potential which are
$O(1)$ in Planck 
units.  Without a fundamental motivation for such a potential, all
arguments about ``naturalness'' and ``fine tuning'' are not very
productive.  We feel, however, that this work represents interesting
progress at a phenomenological level, and might point out promising
directions in which to search for a more fundamental picture.

ACKNOWLEDGMENTS: 
We thank J. Lykken, P. Steinhardt, N. Turok, and B. Nelson for helpful
conversations. 
We  acknowledge support from DOE grant DE-FG03-91ER40674, UC Davis,
and thank the Isaac Newton Institute for hospitality while this
work was completed.

\pagebreak
\pagestyle{empty}

\end{document}